# A Lenient Causal Arrow of Time?

Nathan Argaman

Department of Physics, NRCN, P.O. Box 9001, Be'er Sheva 84190, ISRAEL; argaman@mailaps.org

**Abstract:** One of the basic assumptions underlying Bell's theorem is the causal arrow of time, having to do with temporal order rather than spatial separation. Nonetheless, the physical assumptions regarding causality are seldom studied in this context, and often even go unmentioned, in stark contrast with the many different possible locality conditions which have been studied and elaborated. In the present work, some retrocausal toy-models which reproduce the predictions of Quantum Mechanics for Bell-type correlations are reviewed. It is pointed out that a certain toy-model which is ostensibly superdeterministic – based on denying the free-variable status of some of Quantum Mechanics' input parameters – actually contains within it a complete retrocausal toy-model. Occam's razor thus indicates that the superdeterministic point of view is superfluous. A challenge is to generalize the retrocausal toy-models to a full theory – a reformulation of Quantum Mechanics – in which the standard causal arrow of time would be replaced by a more lenient one: an arrow of time applicable only to macroscopically-available information. In discussing such a reformulation, one finds that many of the perplexing features of Quantum Mechanics could arise naturally, especially in the context of stochastic theories.

**Keywords:** Bell's theorem; the causal arrow of time; retrocausality; superdeterminism; toy-models

**1. Introduction**

Bell's theorem is one of the most profound revelations of modern physics. In the Einstein-Podolsky-Rosen article [1], and in Bell's original proof [2], the discussion is based on notions of locality, but in a later review [3] Bell clarified that the relevant requirement of locality, often called Bell-locality, follows from the assumption of relativistic causality. The original "locality," Bell stated, is in fact simply an abbreviation of "local causality." It is perhaps natural that the original papers, more than the later reviews [3-5], were analyzed meticulously. Moreover, in all of Bell's writings [6,7], and in the overwhelming majority of the accompanying literature, the causal arrow of time is taken for granted, rather than identified as a physical assumption.

What does this assumption mean? Many physicists accept Hume's approach, which defines the concepts of "cause" and "effect" so that a cause always precedes its effect. However, for mathematical models of natural phenomena, it is also natural to take the inputs of a mathematical model as "causes" and the outputs as "effects," which corresponds to regarding the things we can control as causes (see, e.g., [8]). In this context, assuming the causal arrow of time means simply that the model accepts inputs from the past, such as initial conditions, and generates outputs that correspond to later times. Similarly, in such models an external field at a time $t$ affects only the values of variables pertaining to later times. Many of the mathematical models used in physics are precisely of this type, e.g., the standard application of Newton's equations, with initial conditions taken as inputs, or the description of quantum wavefunctions evolving according to Schroedinger's equation between measurements, with Born's rule and the collapse postulate applied at the times of measurements. But some established mathematical models, such as the stationary-action principle of classical mechanics, do not conform to this rule: at the mathematical level, they are in violation of the causal arrow [9] (to avoid confusion, we will use the terms "inputs" and outputs" rather than "causes" and "effects," whether or not a model obeys the causal arrow of time). A reasonable definition of the causal arrow of time for a physical theory is that it is possible to formulate it in terms of a mathematical model with inputs and outputs which conform to the causal arrow stated above.

For deterministic theories, modeled with differential equations, one can specify a concrete solution by choosing as inputs either initial or final conditions (or a combination of both). One can then reserve the identification of causes with inputs (and effects with outputs) only for models which conform to the arrow of time, obtaining consistency with Hume's approach. However, the situation is less clear for stochastic theories, such as QM. In general, it is not guaranteed that a mathematical model which does not conform to the causal arrow of time will have a reformulation which does.

Bell's writings indicate that, while he was interested in stochastic theories, he consistently accepted Hume's approach. For example, in [3] Bell contrasted local causality, which allows for stochastic probabilities, with local determinism, taking for granted that at the mathematical level the past affects the future rather than vice versa. Bell did often mention light cones, but did not pause to explain why the past light cone of an event, rather than its future light cone, is where one may find the inputs affecting it. When considering the possibility that relativistic causality could fail, he discussed a preferred frame of reference in which the causal arrow would still hold [10]. Bell applied relativistic causality to two separated particles which had originated together – the setup of the Einstein-Podolsky-Rosen article – and derived the condition of Bell-locality. Subsequently, many different notions of locality were identified, e.g., "parameter independence" vs. "outcome independence." In fact, a recent article lists no less than eight different statements/definitions of locality [11] (see also [12]). Nonetheless, the logical argumentation requires that we should pay at least as much attention to different definitions of causality as we pay to different notions of locality.

That the causal arrow of time is an essential assumption of Bell's no-go theorem, and hence should be called into question, is a point which was raised repeatedly in the literature [13-16]. When considering theories or models which violate the standard arrow of time, the time-reversal symmetry of microscopic physical theories is often used to argue that introducing any time-asymmetry should be avoided altogether. In the present work, the possibility of introducing such an asymmetry into the theory will be considered, with the aim of reproducing macroscopic phenomena. Specifically, an asymmetric, low-entropy-in-the-past condition will be applied. A directionality of time is expected to result from such asymmetry, but it need not be as strict as the standard causal arrow of time.

For this to work reasonably, one needs to distinguish between microscopic and macroscopic degrees of freedom, and to have the information carried by the macroscopic degrees of freedom constrained by an arrow of time. The microscopic degrees of freedom are to exhibit fluctuations which are affected by inputs from the future, but any attempt at amplifying these fluctuations and bringing them up to the macroscopic level must fail to produce any macroscopic information regarding the future inputs. This corresponds well to known facts concerning quantum fluctuations, the impossibility of using Bell-correlations for signaling, and the disturbance of a quantum system by measurement.

It thus appears that a most promising direction is to pursue retrocausal reformulations of Quantum Mechanics (QM). Reformulations are powerful tools in advancing theoretical physics, e.g., the Lagrangean and Hamiltonian reformulations of Newton's equations played essential roles in the development of QM. In fact, QM had two equivalent formulations to begin with – wave mechanics and matrix mechanics – and some of the most important subsequent advances were based on reformulations, as exemplified by path integrals. Additional examples include Bohmian mechanics [17], which motivated Bell in his original research on no-go theorems [18], and its stochastic version – diffusing particles guided by the quantum wavefunction, which was suggested by Bohm himself [19] and further developed by others [20]. Improving our understanding of QM appears to require such a radical retrocausal reformulation of the fundamental theory. This could have ramifications in contexts such as quantum computation and/or quantum gravity.

Notice that this approach is complementary to, but quite distinct from the experimental approach, which is by and large the main activity following from Bell's theorem, e.g., [21]. The empirical adequacy of QM has by now been abundantly supported, but can never reach the certitude of a mathematical theorem. The proven theorem states that the predictions of QM (*i.e.*, the probabilities for the output parameters, given the input parameters) cannot be reproduced by a model or theory conforming to the condition of Bell-locality. Thus, our discussion will relate to models or theories which violate this mathematical condition, by allowing retrocausality. Questions regarding the real state of the system, such as those which concerned Einstein [1], and which arise naturally when the theory is compared with experimental procedures, do not arise in this context of model-construction. In particular, the issue of counterfactual definiteness [22] is not relevant, because the discussion does not refer to the question whether or not the real system has a definite property, but instead to the much simpler question of whether or not the model or theory has a prediction for that property. The relationship between reality and the parameters and predictions of the theory is adequately handled by standard QM (and its interpretations), and need not be addressed when considering reformulations.

As a first step in this direction, it is appropriate to discuss retrocausal toy-models, *i.e.*, mathematical models which reproduce the predictions of QM for the specific case considered in Bell's theorem. Two of the toy-models available will be reviewed. The first [16] was presented as a mathematical formulation of the retrocausal ideas expressed, e.g., by Cramer [14]. The second [23] was originally presented in a somewhat different context, associated with superdeterminism, *i.e.*, the denial of the free-variable status of the inputs of QM, which is generally associated with the free will of the experimenters. It will be argued that while the latter model has a distinct technical advantage, the former presentation is more relevant as a basis for a discussion of future scientific theories.

The two retrocausal toy-models discussed share the undesirable feature of QM known as the measurement problem. On the other hand, they differ in that the "dynamics" in one is stochastic, while in the other, deterministic. Can they be generalized to encompass all quantum phenomena, in a way which would allow an understanding of both nonlocality and quantum measurements? This possibility will be qualitatively discussed, assuming the stochastic option for the dynamics. In this context, it seems that many of the other mysterious aspects of QM, such as the exponentially large size of the Hilbert space required to describe $n$ particles, and the dynamics involving unitary evolution punctuated by collapse, follow naturally. This discussion could be compared and contrasted with the questions regarding the relative sizes of ontic and epistemic spaces in the context of the ontological models framework, which is based on assuming strict causality. The difficulties that that framework faces [24] serve as further motivation for considering the retrocausal alternative.

The discussion of toy-models is the subject of Section 2, the pathway towards a general retrocausal reformulation of QM is discussed in Section 3, and conclusions are provided in Section 4.

## 2. Retrocausal toy-models

Bell's theorem concerns entangled particles, e.g., pairs of distant photons with polarizations entangled in a singlet state, as in some of the most remarkable early experiments [25] (originally, pairs of spin-half particles were considered). The predictions of QM for polarization measurements on the constituent photons of such a pair are given by the probabilities

$$p_{a,b}(A,B) = \frac{1}{4}[1 + A B \cos(2a - 2b)], \qquad (1)$$

where $a$ and $b$ are angles, defined modulo $\pi$, specifying the orientations of the beam-splitting polarizers involved in the measurements, and $A, B = \pm 1$ represent the results of the measurements. $A = 1$ represents the first photon having a polarization along $a$, $A = -1$ represents a perpendicular polarization, and the polarization of the second photon with respect to the orientation $b$ is similarly represented by $B$. In principle, even when one restricts QM to a description of the polarizations of a pair of photons, one has an additional input variable $c$ specifying how the photons are prepared, and thus determining the initial wavefunction. However, in the present work we will only refer to the singlet state, and thus $c$ is a constant, and can be dropped. The probabilities (1) are "local" in the no-signaling sense, *i.e.*, the marginal probabilities are independent of the remote inputs: $p_{a,b}(A) \equiv \sum_B p_{a,b}(A,B) = \frac{1}{2}$ is independent of $b$, and similarly $p_{a,b}(B)$ is independent of $a$. However, they are "nonlocal" in the sense of Bell, *i.e.*, one cannot write $p_{a,b}(A,B)$ as a product of two separate factors, $p_{a,\lambda}(A)$ and $p_{b,\lambda}(B)$, where $\lambda$ is an additional parameter (or set of variables) describing the "state" the particles had in the past, which is taken to be independent of the inputs $a$ and $b$.

*A simplistic toy-model*

The idea of retrocausality is to recognize this last restriction as a physical assumption – the causal arrow of time – which may be inappropriate for a microscopic theory. If one allows $\lambda$ to depend on $a$ and $b$, the difficulty is resolved, as demonstrated by the following simplistic model [16]:

$$p_{a,b}(\lambda) = \frac{1}{4}\left[\delta(\lambda - a) + \delta\left(\lambda - a - \frac{\pi}{2}\right) + \delta(\lambda - b) + \delta\left(\lambda - b - \frac{\pi}{2}\right)\right], \quad (2)$$

where $\lambda$ is an additional angle, defined modulo $\pi$, which represents the initial polarization of the photons. In this model, the predictions for the polarization measurements follow from the standard Malus' law,

$$p_{a,\lambda}(A) = \begin{cases} \cos^2(\lambda - a) & A = 1 \\ \sin^2(\lambda - a) & A = -1 \end{cases}, \quad (3)$$

and similarly for $p_{b,\lambda}(B)$. Combining these using

$$p_{a,b}(A,B) = \int d\lambda \; p_{a,b}(\lambda) p_{a,\lambda}(A) p_{b,\lambda}(B) \quad (4)$$

reproduces the predictions of QM, Eq. (1). A toy-model which approaches this simplistic one in the appropriate limit ($\gamma \to 0$) is discussed in [26].

An attractive feature of this model is that one can consider what would happen if the value of $\lambda$ were measured at the source. In the experiments [25], the singlet pair of photons was emitted by an atomic $(J = 0) \to (J = 1) \to (J = 0)$ cascade. One can envision measuring the orientation of the angular momentum of the atom during the brief time it is in the intermediate state of the cascade, thereby inferring the initial polarization of the photons. In order to perform such a measurement, one must specify the direction in which the angular momentum is measured, e.g., whether it is a measurement of $\hat{J}_x$ or $\hat{J}_y$ (the results of this measurement and those of each one of the later photon-polarization measurements would be correlated per the predictions of QM). Clearly, this would constitute a "which path" measurement [27], and would disturb the system in a manner which ruins the entanglement between the two photons.

*Hall's toy-model*

According to (2), the variable $\lambda$ carries much information regarding the values of $a$ and $b$, or at least one of them. It is of interest to note that this "flow of information from the future to the past" is much more limited in other toy-models. A very efficient model, in this respect, was given in [23]. The variant of it pertaining to photons, rather than spin-half particles, is:

$$p_{a,b}(\lambda) = \frac{1}{\pi} \frac{1 + \acute{A}\acute{B}\cos(2a - 2b)}{1 + \acute{A}\acute{B}(1 - z)}, \quad (5)$$

where $\acute{A} = \text{sign}[\cos(2a - 2\lambda)]$, $\acute{B} = \text{sign}[\cos(2b - 2\lambda)]$ and $z = \frac{2}{\pi}|2a - 2b|$ are abbreviations (the denominator never vanishes, because $\acute{A}\acute{B} = 1$ when $z = 0$ and $\acute{A}\acute{B} = -1$ when $z = 2$). This model is deterministic in the sense that $a$ and $\lambda$ determine one possible value for $A$,

$$p_{a,\lambda}(A) = \delta_{A,\acute{A}}, \quad (6)$$

and similarly for $B$ and $p_{b,\lambda}(B)$. Again, using Eq. (4) to combine Eqs. (5) and (6) trivially reproduces the predictions of QM, Eq. (1).

In the case of Eq. (5), the "past" variable $\lambda$ carries very little information on $a$ and $b$, less than 0.07 bits [28]. This can justifiably be seen as a definite advantage over the simplistic model of Eq. (2). On the other hand, Refs. [23] and [28] can be criticized for not discussing retrocausality explicitly (although its relevance was briefly acknowledged by the author in [29]). In fact, the possibility that the variable $\lambda$ depends on the inputs $a$ and $b$, embodied in (5), is presented in [23] in a manner that does not imply a violation of the causal arrow of time.

*Criticism of the superdeterministic approach*

How could this come about? As described in the introduction, Bell emphasized the locality assumption, $p_{a,b,\lambda}(A,B) = p_{a,\lambda}(A)\, p_{b,\lambda}(B)$. In 1976 he was criticized by Shimony *et al.* [30] for not emphasizing the measurement-independence assumption,

$$p_{a,b}(\lambda) = p(\lambda) \; , \tag{7}$$

as well. Bell replied that he *had* (belatedly) made explicit the assumption that $a$ and $b$ were free variables, and that "this means that the values of such variables have implications only in their future light cones" [31]. He also emphasized that his work should not be understood as a philosophical discussion concerning the real world, but as an analysis of the kinds of mathematical models or theories which may be applicable. Obviously, the mathematical notion of free variables, which pertains to such theories, is not specific to situations in which the past and/or the future are relevant. Thus, the measurement-independence assumption relies on two separate assumptions:

    (i) $a$ and $b$ are free variables;
    (ii) The causal arrow of time, with $\lambda$ associated with a time earlier than that of $a$ and $b$.

The first implies that $a$ and $b$ are independent of $\lambda$, and the second that $\lambda$ is independent of $a$ and $b$, giving a full justification of (7). However, saying that they are independent of each other might be misleading, because the word "independent" is used here with different mathematical meanings. In particular, mutual statistical independence is applicable only if the free-variable status of $a$ and $b$ is revoked, and they are replaced by random variables.

If indeed $a$ and $b$ are treated as random variables, like $\lambda$, then the probability distribution $p_{a,b}(\lambda)$ is replaced by a conditional probability, $p(\lambda|a,b)$ (Bell's notation, $\{\lambda|a,b\}$, could refer to either of these). The measurement independence condition (7) then reads $p(\lambda|a,b) = p(\lambda)$, and it no longer follows from the arrow of time alone – a conditional dependence of $\lambda$ on $a$ and $b$ could also arise from a forwards-in-time dependence of $a$ and/or $b$ on $\lambda$, or from a common cause in the past. The latter is indeed the possibility considered in [30], which describes a conspiracy involving a person who has "concocted" a list of correlation data, an apparatus manufacturer, and the secretaries of two physicists who are to perform the experiments. When the $j$th measurement is to be performed, each of the secretaries whispers the pre-listed setting to the corresponding physicist, who sets his apparatus accordingly, and the result registered by each apparatus is pre-programmed to correspond to the concocted list, rather than to an actual measurement. In this manner, any correlations can result, including those of (1), with no violations of local causality, but this is achieved conspiratorially.

The agreement of empirical observations with QM, whether $a$ and $b$ are selected at random or are determined by any other arrangement (say, a double-blind experimental procedure), provides strong evidence that it is appropriate to treat the orientations as free variables, and to avoid serious discussion of the possibility that they could be predetermined. Indeed, it was argued from the outset (in the concluding paragraphs of [30]) that the "enterprise of discovering the laws of nature" by "scientific experimentation" necessarily involves the assumption that "hidden conspiracies of this sort do not occur." Accordingly, the purpose of the discussion was not to cast doubt on the free-variable status of $a$ and $b$, but merely to point out that a complete statement of Bell's theorem must include a reference to this status, assumption (i) above. Indeed, from that point on, Bell emphasized that the proof of his theorem assumes this, and engaged in brief discussions of the possibility of "superdeterministic" theories, in which there are no such free variables [4,5,31], concluding that even in such cases pseudorandom variables which are "sufficiently free for the purpose at hand" should be available. Subsequently, violations of measurement independence, Eq. (7), were often perceived as equivalent to violations of assumption (i), involving free variables or "free will," and the similarly critical assumption (ii), involving the arrow of time, very often remained unmentioned.

As described above, the model of Eqs. (5) and (6) is presented in [23] in this manner – the violation of measurement independence is emphasized, the "free will" issue is discussed, and the role of the causal arrow of time is not. Neither are the unscientific/conspiratorial concern and the notion of pseudorandom variables. In fact, in a more complete presentation of the model, the author finds it necessary to add a variable $\mu$ associated with the overlap of the backwards lightcones of $a$ and $b$, and determining them (Section 5.1 of [28]). For an explicit version of the toy-model, it is suggested that $\mu$ simply consists of the values of $a$ and $b$, determining them in a decidedly artificial manner. In addition, it was found necessary to invoke a correlation-does-not-imply-causation argument to explain how the ostensibly retrocausal dependence in Eq. (5) is consistent with causality. When considering Eqs. (5) and (6) as a retrocausal rather than a free-variable-status-denying model, these additional steps are superfluous. Thus, one may add Occam's razor as a further argument to prefer the former over the latter, *i.e.*, to interpret violations of measurement independence as violations of (ii) rather than of (i). The details given above are in the context of Refs. [23] and [28], but on the basis of the arguments given, the Occam's razor argument is expected to hold quite generally for superdeterministic models. Nevertheless, such models are often regarded as the leading alternative to standard QM, by both experimentalists (e.g., [32]) and leading theorists (e.g., [33]).

## 3. Toward a general retrocausal theory

Can the retrocausal toy-models discussed above be generalized to a full retrocausal theory of all quantum phenomena, including a resolution of the measurement problem? The development of such a theory would be revolutionary, as it would go beyond all previous developments in the foundations of QM: path integrals [34], Bohmian mechanics [17], histories approaches (e.g., [35]), stochastic mechanics [20], stochastic quantization [36], etc. In contemplating such a development, promoters of retrocausation often advocate a fully time-reversal-symmetric approach [15]. However, to bring microscopic theories in line with macroscopic phenomena, it is common in other contexts to break this symmetry by invoking a low-entropy condition in the remote past. It is therefore of interest to consider how a "fixed past" initial condition would affect different types of theories, as sketched in Figure 1 (see also [37]).

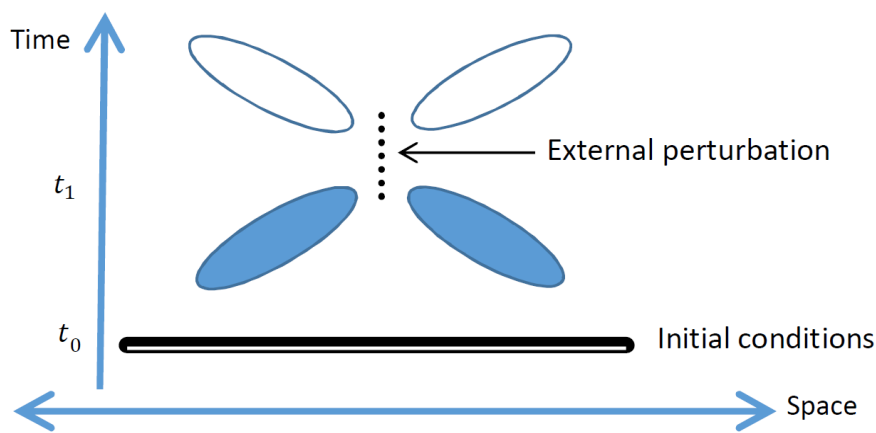

**Figure 1.** Sketch of a spacetime region permeated by fluctuating fields, with an external perturbation applied at times near $t_1$, and with the field configuration at an early time, $t_0$, fixed as initial conditions. If the fields are described by a deterministic theory, the field configurations before $t_1$ are unaffected by the external perturbation; in contrast, for stochastic theories the probability distribution of the fields at times between $t_0$ and $t_1$, indicated by the filled ovals, may depend on the perturbation. Nevertheless, the firing rate of a "detector" placed at or near these ovals must not depend on the perturbation (or else the no-signaling condition would be violated).

Consider a theoretical description of some degrees of freedom in spacetime, such as the modes of an electromagnetic field, and consider an external perturbation (possibly an oscillating dipole) which affects the dynamics of these degrees of freedom for a limited time, beginning at a time $t_1$. Furthermore, consider the possibility that the values of all the degrees of freedom described by the theory are given at a time $t_0$ in the past (if the dynamics is described by a second-order differential equation in time, the time derivatives of these values are also considered here as given). For a theory with deterministic dynamics, full specification of the field configuration at time $t_0$ would determine the field configuration for all times up to $t_1$, at which the external perturbation is applied. However, if the theory describing the dynamics of the fields is stochastic, this is no longer necessarily true and the probabilities for the fields at times before $t_1$ may be retrocausally affected by the external perturbation. In the case in which the dynamics allows for only small fluctuations around the classical behavior, such as the quantum vacuum fluctuations around a zero-field solution, one may expect a weaker form of causality to still hold.

If the theory in question can also model the process of measurement, it must be capable of describing the inner workings of a detector, which could be placed within the regions marked by the open or the filled ovals in the sketch (times later than or earlier than $t_1$, respectively). The detector itself would consist of additional stochastic degrees of freedom, and would presumably have an initial condition corresponding to a metastable state. Fluctuating out of the metastable state would correspond to a "click" in the detector, and the probability for such a fluctuation would depend on the fluctuations on the field to which the detector is coupled. Dissipative aspects of the detector could be modeled by coupling it to a "bath" of many "thermalized" degrees of freedom, subject to appropriate initial conditions of their own, in analogy to the Caldeira-Leggett description of quantum dissipation [38]. An "ideal" detector would correspond to the case for which the "click" is sufficiently dissipative to be "irreversible," so that coupling of further degrees of freedom to the detector would allow for copying or "cloning" of the information regarding whether or not a "click" has occurred. This would stand in contrast to the no-cloning condition which is expected to hold in general for stochastic theories (see, e.g., [39]).

If the detector operates at times after $t_1$ (the empty ovals), it is natural for the probability of detection to depend on the details of the external perturbation. However, if the detector operates at times earlier than $t_1$, it is necessary to require that the corresponding probability is independent of the external perturbation, in order to avoid the possibility of signaling to the past. In other words, although the fluctuating fields are subject to rules which allow for retrocausality, the external perturbation and the detection events must be related in a manner which is subject to the causal arrow of time. As this causal arrow applies only to the "macroscopically available" or "clone-able" information, and does not affect the microscopic "hidden" degrees of freedom, it is perhaps appropriate to characterize it as "lenient." The fixed initial conditions are associated with low entropy, and therefore may be expected to break the symmetry of the theory in just the manner required to meet such a lenient causality requirement.

Note that we are here considering the possibility that an "agent" which is external to the theory may control the inputs and may use the information provided by the measured outputs, and that such an agent is subject to the arrow of time as well. For example, the agent may decide whether or not to apply the external perturbation according to whether or not a detection event has occurred at an earlier time. Signaling into the past must thus indeed be strictly impossible, as it would allow construction of causal loops – the well-known inconsistency arguments of the grandfather paradox (a.k.a. the bilking argument). However, these arguments involve only the information accessible at the macroscopic level, so they precisely allow the type of lenient causality described here, in which the microscopic degrees of freedom are affected retrocausally [15].

A theory of this type, if developed, would provide a natural explanation for many of the perplexing features of QM. The description above bears similarities to the analysis of quantum measurement by environment-induced superselection, or einselection [40]. It would be natural to supplement it by an alternative description which would represent only the macroscopically-available information regarding a certain subset of the degrees of freedom up to a time $t$. As the external perturbations applied at later times to these degrees of freedom are to be treated as unknown, this mathematical description would have to represent a large number of possibilities, exponentially large in the number of degrees of freedom involved. Furthermore, by definition it would have to evolve in an information-preserving manner as $t$ is changed, except for the moments at which there is a change in the macroscopically-available information. This corresponds precisely to the evolution of quantum wavefunctions, which form exponentially large Hilbert spaces, and exhibit unitary evolution punctuated by "collapse" events (see also [41]).

For a specific retrocuasal theory of the type considered here, it is expected that a direct generalization of the deduction of the lenient causality condition would lead to the slightly stronger condition of "information causality" [42]. This latter condition was put forward as a physical principle within an axiomatic approach, *i.e.*, with the hope that all of the features of QM could be deduced from it. This was partly successful, as it was demonstrated that information causality implies Tsirelson's bound [43], a generalization of the Bell (or CHSH) inequality which holds in the quantum realm. In this sense, it is expected that time-asymmetric retrocausal theories of the type considered here would, through the mathematical arguments of [42], provide an explanation for the fact that all quantum phenomena obey Tsirelson's inequality. Note that information causality is here to be deduced rather than assumed, and thus the fact that not all aspects of QM can be generated by it does not lead to any objections in the present context.

## 4. Summary and Discussion

The present work, like several other presentations at the EmQM17 David Bohm centennial conference [44–47], advocates the relaxation of the arrow-of-time assumption of Bell's theorem. In the first part, the role this assumption plays in the proof of the theorem was considered, and contrasted with the free-variable assumption, which is associated with the free will of the experimenters. Concrete toy-models which violate these assumptions were discussed. In the second part, an admittedly speculative discussion of the possibility of developing a retrocausal reformulation of QM which would describe all quantum phenomena in spacetime (rather than a Hilbert space), and would be free of the measurement problem, was given.

In the proof of Bell's theorem, the arrow-of-time assumption enters together with the free-variable status of the measurement settings, leading to the mathematical condition of measurement independence [Eq. (7) above]. Unfortunately, the arrow of time is usually taken for granted, rather than identified as a physical assumption, and relaxation of the measurement-independence condition is then associated with superdeterminism, *i.e.*, denial of the free-variable status of the settings, rather than with retrocausality.

It was pointed out from the outset that for superdeterminism to provide a resolution of the difficulty exposed by Bell's theorem, one must assume that the measurement settings are produced in a conspiratorial manner, one which would undermine the scientific method [30]. In contrast, accepting violations of the mathematical causal-arrow-of-time condition was necessary already in the context of the stationary-action principle of classical mechanics. Thus, while both face the difficulty of overcoming the prejudices we have developed based on our experiences in the macroscopic world, there is a clear preference for the latter over the former. Nevertheless, the superdeterministic approach has received much attention recently, and several concrete examples of superdeterministic toy-models have been put forward. This may be due to the argument concerning the grandfather paradox, but such paradoxes are easily avoided if one takes retrocausality to affect only hidden variables [15], such as the which-path variables of standard QM [27] (those whose measurement precludes the observation of interference effects).

As a further argument against superdeterminism, one of the available toy-models [23] has been considered, and it was shown that it consists of a complete retrocausal toy-model, to which additional steps of argumentation have been added in order to transform it into a superdeterministic toy-model. Occam's razor thus rules against the superdeterministic approach, and for the retrocausal interpretation of such toy-models. When interpreted in this manner, this particular toy-model of Ref.

[23] has a distinct advantage over an earlier retrocausal toy-model [16], in that the microscopic degrees of freedom carry a very limited amount of information into the past.

Developing a general retrocausal theory of quantum phenomena, one that would be free of the measurement problem and not limited to the scope of a toy-model, is a grand challenge of quantum foundations. Whereas it is appropriate to discuss fully time-symmetric theories in this context [15], the possibility of breaking the symmetry by imposing fixed boundary conditions in the past was considered above. A theory with fully fixed initial conditions and deterministic dynamical rules cannot exhibit retrocausality of the type required by Bell's theorem. In contrast, a theory with time-symmetric stochastic dynamical rules would have its symmetry broken by the imposition of initial conditions, in a manner which may lead to a "lenient" arrow of time of the type observed macroscopically – an arrow of time applicable only to macroscopically available information.

Such a stochastic time-asymmetric approach is expected to enjoy two further advantages: (a) Constructing the corresponding "epistemic" state of knowledge, *i.e.*, a mathematical representation for the macroscopically-available information up to a time $t$, would necessarily result in states which are exponentially complex for many degrees of freedom. These states would "evolve" with $t$ in an information-preserving manner, except at the times of measurements, at which additional information becomes available. The correspondence to the complexity of quantum wavefunctions and their unitary/collapse evolution is clear. (b) The spatiotemporal "flow" of entropy/information in such theories is expected to lead to the information causality condition, and thus to Tsirelson's bound.

In closing, it is appropriate to quote the concluding paragraph of Bell's last review of his theorem [5]:

> The unlikelihood of finding a sharp answer to this question [the measurement problem] reminds me of the relation of thermodynamics to fundamental theory. The more closely one looks at the fundamental laws of physics the less one sees of the laws of thermodynamics. The increase of entropy emerges only for large complicated systems, in an approximation depending on "largeness" and "complexity." Could it be that causal structure emerges only in something like a "thermodynamic" approximation, where the notions "measurement" and "external field" become legitimate approximations? Maybe that is part of the story, but I do not think it can be all. Local commutativity does not for me have a thermodynamic air about it. It is a challenge now to couple it with sharp internal concepts, rather than vague external ones. …

Developing a fundamental retrocausal stochastic theory, may resolve the issue, as the condition of local commutativity need only apply to the corresponding epistemic states, representing the macroscopically-available information. It is expected that the process of mathematically constructing such epistemic states would provide the "sharp internal concepts" required by Bell to meet this challenge.

**Acknowledgments:** This work was presented at the EmQM17 David Bohm centennial conference, and its publication was supported by the Fetzer Franklin Fund of the John E. Fetzer Memorial Trust.

**Conflicts of Interest:** The author declares no conflict of interest.